\documentclass[twocolumn,prl,aps]{revtex4-1}
\usepackage[T1]{fontenc}
\usepackage[latin9]{inputenc}
\usepackage{color}
\usepackage{bm}
\usepackage{amsmath}
\usepackage{amssymb}
\usepackage{graphicx}
\usepackage{braket}
\usepackage{esint}
\PassOptionsToPackage{normalem}{ulem}
\usepackage{ulem}
\usepackage[unicode=true,bookmarks=false,breaklinks=false,pdfborder={0 0 1},backref=false,colorlinks=true]{hyperref}
\hypersetup{pdfborderstyle=}

\makeatletter

\usepackage{color}
\usepackage{bm}
\usepackage{bbold}
\usepackage{soul}
\usepackage{epsfig}
\usepackage{verbatim}
\usepackage{array}
\newcommand{\be}{\begin{equation}}
\newcommand{\ee}{\end{equation}}
\newcommand{\bea}{\begin{eqnarray}}
\newcommand{\eea}{\end{eqnarray}}

\makeatother

\begin{document}
\global\long\def\sp#1#2{\langle#1|#2\rangle}%
\global\long\def\abs#1{\left|#1\right|}%
\global\long\def\avg#1{\langle#1\rangle}%
\global\long\def\ket#1{|#1\rangle}%
\global\long\def\bra#1{\langle#1|}%

\title{Driven Dissipative Majorana Dark Spaces}
\author{Matthias Gau,$^{1,2}$ Reinhold Egger,$^{1}$ Alex Zazunov,$^{1}$ and Yuval Gefen$^{2}$}
\affiliation{$^{1}$~Institut f\"ur Theoretische Physik, Heinrich-Heine-Universit\"at,
D-40225 D\"usseldorf, Germany\\
$^{2}$~Department of Condensed Matter Physics, Weizmann Institute, Rehovot, Israel 
}
\date{\today}
\begin{abstract}
Pure quantum states can be stabilized in open quantum systems subject to external driving forces and dissipation by environmental modes.  
We show that driven dissipative (DD) Majorana devices offer key advantages for stabilizing degenerate state manifolds (`dark spaces')
and for manipulating states in dark spaces, both with respect to native (non-DD) Majorana devices and to
DD platforms with topologically trivial building blocks.  For two tunnel-coupled Majorana boxes,  using otherwise 
only standard hardware elements (e.g., a noisy electromagnetic environment and quantum dots with driven tunnel links), 
we propose a dark qubit encoding. We anticipate exceptionally high fault tolerance levels due to a conspiracy of DD-based autonomous error correction and topology.
\end{abstract}
\maketitle

\emph{Introduction.---}Open quantum systems may be stabilized  in a pure quantum state for arbitrarily long times by the interaction with an external driving field and a dissipative environment \cite{Plenio1999,Beige2000,Plenio2002,Diehl2008,Kraus2008,Diehl2010,Diehl2011,Bardyn2013,Zanardi2014,Albert2014,Jacobs2014,Albert2016,Goldman2016}. 
Such DD-stabilized dark states are eigenstates of the Lindbladian dissipator with zero eigenvalue when the system dynamics can be described by a Lindblad equation   \cite{Lindblad1976,Lindblad1983,Gardiner2004,Breuer2006,Wiseman2010}.
The latter is the most general Markovian master equation preserving the trace and semi-positiveness of the density matrix.
Using trapped ions or superconducting qubits, DD-stabilized dark states have recently been implemented experimentally \cite{Geerlings2013,Lu2017,Touzard2018,Barreiro2011,Schindler2013,Shankar2013,Leghtas2013,Reiter2016,Liu2016}.
For a stabilized manifold of multiple degenerate dark states (a dark space) \cite{Touzard2018,Iemini2015,Iemini2016,Santos2020}, a robust quantum memory platform can be envisioned. 
Moreover, once states within a dark space can also be manipulated in a protected way, fault-tolerant quantum computing schemes without active feedback may become a viable option, see Refs.~\cite{Verstraete2009,Fujii2014,Terhal2015,Kerckhoff2010,Murch2012,Kapit2015,Kapit2016,Herold2017} for related work. At present, experimental studies of autonomous error correction in a  DD qubit~\cite{Leghtas2013,Liu2016,Reiter2017,Puri2019} report fidelities below 90$\%$ for state stabilization and significantly lower fidelity for gate operations.

\begin{figure}[t]
\begin{centering}
\includegraphics[width=\columnwidth]{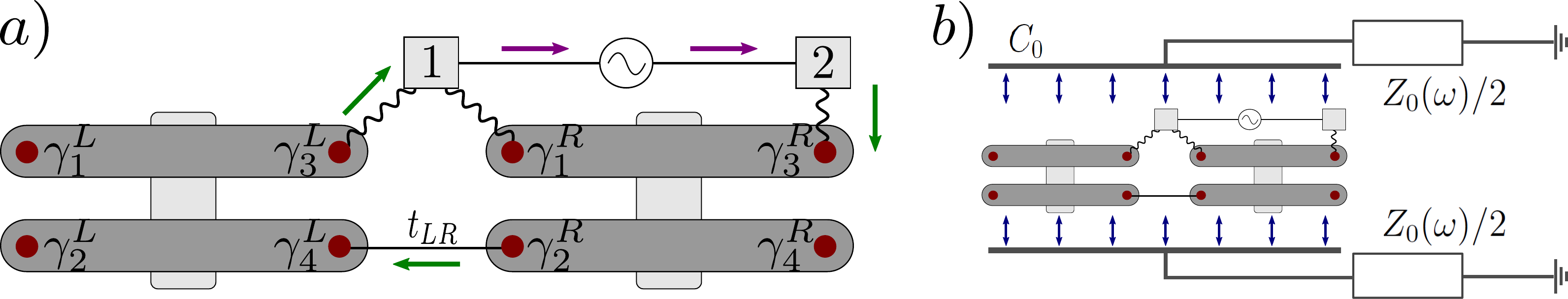}
\end{centering}
\caption{Driven dissipative Majorana device with two tunnel-coupled Majorana boxes. (a) Each island harbors four Majorana states ($\gamma_{\nu}^{\kappa=L/R}$, red dots).
The latter are shown as end states of topological hybrid nanowires (horizontal bars), which are connected by a superconducting bridge (vertical bar) to form a floating mesoscopic island \cite{Plugge2017,Karzig2017}.
Two quantum dots (boxes 1 and 2) are connected by tunnel contacts (wavy lines) to Majoranas.
A driven tunnel link (solid line) connects both dots.
The colored arrows illustrate the pumping-cotunneling cycle
explained in the main text, stressing the nonlocal character of the
processes involved.  The shown trajectory amounts to the action of the
operator $Z_LY_R$ on the Majorana state, see Eq.~\eqref{PauliOp}. 
(b) Electromagnetic fluctuations of the surrounding electric circuit, modeled by the capacitance $C_0$ and the impedance $Z_0(\omega)$, 
cause inelastic tunneling processes.  }
\label{fig1}
\end{figure}

In this Letter, we show that devices harboring Majorana bound states (MBSs) \cite{Alicea2012,Leijnse2012,Beenakker2013,Sarma2015,Aguado2017,Lutchyn2018,Zhang2019a,Liu2018b,Zhang2018b,Wang2018b,Sajadi2018,Ghatak2018,Murani2019} provide a particularly attractive platform for the DD stabilization of degenerate dark spaces and for manipulating states in such spaces. The use of topologically protected building blocks for stabilizing DD dark spaces offers several key
advantages for quantum state stabilization and manipulation protocols when compared to DD schemes with topologically trivial building blocks or to topological platforms without DD protection: 
(i)  Majorana-based dark spaces benefit from both  topological protection and DD-based autonomous error correction capabilities. 
In particular, the reduced intrinsic noise levels expected from the topological protection help to avoid unwanted residual dissipation effects within the dark space manifold \cite{Facchi2000}. 
(ii)  As indicated in Fig.~\ref{fig1}, our DD protocols exploit unidirectional cotunneling processes 
between pairs of quantum dots (QDs), which are connected by tunnel contacts to  MBSs and by a driven tunnel link to each other.
Dissipation here originates in a natural way from the electromagnetic environment.  A weak driving field serves to pump electrons from QD 1 to the energetically
high-lying QD 2 in Fig.~\ref{fig1}, and the electron transfer from QD $2\to 1$ then proceeds by inelastic cotunneling.  
By choosing pre-designated tunnel couplings, this pumping-cotunneling cycle allows one to  engineer at will jump operators acting on the Majorana state. 
Once a working Majorana platform becomes available, only standard hardware elements are needed to realize the proposed DD Majorana setups.  
(iii) The MBS-based dark space stabilization is very robust with respect to variations of stabilization parameters. 
Deviations of up to $\approx 10\%$ in these gate-tunable parameters are tolerable while retaining almost perfect fidelity. 
This remarkable degree of robustness is due to the spatial disentanglement of drive and dissipation processes, see Fig.~\ref{fig1}, which in turn is connected to the nonlocality of the MBS system.  
 (iv) State preparation and manipulation protocols within the dark space
can be implemented in a flexible and rather simple manner.

We here illustrate these points for a device with two tunnel-coupled Majorana boxes \cite{Plugge2017,Karzig2017} operated under Coulomb blockade conditions, see Fig.~\ref{fig1} for a schematic sketch. Each box harbors  four MBSs.
Our main results are as follows.  We first show that over a wide parameter regime, the dynamics in the Majorana sector is governed by a Lindblad equation, Eq.~\eqref{LindbladMBQ} below, which includes
a Hamiltonian  describing the unitary part of the time evolution and a Lindbladian dissipator  responsible for the dissipative dynamics.  Second, we demonstrate that 
this system can be engineered to support a multi-dimensional degenerate dark space.  Third, to ensure that a generic initial state evolves towards
a designated pure state within the dark space, one has to adiabatically break the degeneracy of the dark space
during intermediate stages of the protocol. We view this as a paradigmatic protocol (applicable to even more complex systems) for the preparation and manipulation of a state within a degenerate dark space, where we also 
explain how to optimize the speed of approach and
the fidelity. Finally, we show how to manipulate states within the dark space.  
Our results are illustrated for a dark space that is equivalent to a fault-tolerant dark qubit. We note that in networks of coupled 
dark qubits, the active error correction required in Majorana surface code proposals \cite{Terhal2012,Vijay2015,Landau2016,Plugge2016,Litinski2017,Wille2019} could become obsolete. 
Detailed derivations and specific state stabilization protocols are described in Ref.~\cite{PRB}.

\emph{Model.---}Consider the architecture in  Fig.~\ref{fig1}, where two Coulomb-blockaded topological superconductor islands ($\kappa=L/R$ for the left/right island) harbor in total eight MBSs with operators $\gamma_{\nu}^\kappa=(\gamma_{\nu}^\kappa)^\dagger$, where $\nu=1,\ldots,4$. They obey the anticommutation relations $\{\gamma_\nu^\kappa,\gamma_{\nu'}^{\kappa'}\}=2\delta_{\nu\nu'}\delta_{\kappa\kappa'}$.  We here assume that all MBSs are sufficiently far away from each other to represent zero-energy states. 
In addition, the relevant energy scales should be below the pairing gap $\Delta$ such that above-gap quasiparticles can be neglected. Each mesoscopic island in Fig.~\ref{fig1} has a large (and, for simplicity, equal) charging energy $E_C$ and is operated under Coulomb valley conditions.  
At temperatures  $T\ll E_C$, charge on the island is then quantized on time scales $\delta t>1/E_C$, implying a parity constraint for the Majorana sector of each box, $\gamma^\kappa_1\gamma^\kappa_2\gamma^\kappa_3\gamma^\kappa_4=\pm 1$ \cite{Karzig2017}.
Since charge is gapped out, the remaining twofold degeneracy of the ground state of each island corresponds to the presence of two Majorana qubits \cite{Karzig2017}.  The respective Pauli operators  are \cite{Beri2012}
\begin{equation}
    X_\kappa = i\gamma_1^\kappa\gamma_3^\kappa,\quad Y_\kappa = i\gamma_3^\kappa\gamma_2^\kappa,\quad 
    Z_\kappa = i\gamma_1^\kappa\gamma_2^\kappa.\label{PauliOp}
\end{equation}
This nonlocal representation allows one to access all Pauli operators 
through electron cotunneling processes between pairs of tunnel-coupled QDs \cite{Landau2016,Plugge2016,Plugge2017,Karzig2017}. We
also need  a phase-coherent tunnel link between the boxes,  $H_{LR} = i t_{LR} \gamma_4^L \gamma_2^R,$ with  real-valued  $t_{LR}>0$.

The single-level QDs in Fig.~\ref{fig1}, 
with electron annihilation operator $d_{j}$ for the $j$th QD, are described by the dot Hamiltonian $H_{\rm d}=\sum_{j=1,2}\epsilon_j d_j^\dagger d_j^{}$, where the level energies $\epsilon_1<\epsilon_2$ should satisfy  $|\epsilon_j|\ll E_C,\Delta$.  The QDs are connected by a driven tunnel link, which we model by  $H_{\rm drive}(t)=2A\cos(\omega_0 t) d_1^\dagger d_2^{}+{\rm h.c.}$ \cite{Platero2004} with drive amplitude $A$.
The driving frequency $\omega_0$ is tuned in resonance with the transition energy between the QD levels, $\omega_0=\epsilon_2-\epsilon_1$.  
Since the Majorana boxes are operated under Coulomb valley conditions, the total occupancy 
$N_{\rm d}=\sum_j d_j^\dagger d_j^{}$ of the QDs is also conserved on time scales $\delta t>1/E_C$ \cite{Romito2014}.  
We study the case $N_{\rm d}=1$, where a single electron is shared by both QDs.
Finally, inelastic tunneling processes connecting QDs with the respective MBSs in Fig.~\ref{fig1} are modeled by 
\begin{equation}\label{tunnelHam}
    H_{\rm tun}=t_0 \sum_{j,\nu,\kappa} \lambda_{j,\nu\kappa}^{}e^{-i \phi_\kappa} e^{i\theta_j} d^\dagger_j  
     \gamma_{\nu}^\kappa  +{\rm h.c.},
\end{equation}
where the $e^{i\phi_\kappa}$  ($e^{-i\phi_\kappa}$) factors in
Eq.~\eqref{tunnelHam} ensure that an electron charge is added to (subtracted from) the respective island in a tunneling process \cite{Devoret1990,Girvin1990,Fu2010}.
With the overall energy scale $t_0\ll E_C$, the complex-valued 
parameters $\lambda_{j,\nu\kappa}$ with $|\lambda_{j,\nu\kappa}|\leq 1$ encode the transparency of the tunnel contact between
$d_j$ and  $\gamma_\nu^{\kappa}$ \cite{Zazunov2016}. For the setup in Fig.~\ref{fig1}, the only non-zero parameters are 
 $\lambda_{1,3L}$, $\lambda_{1,1R}$ and $\lambda_{2,3R}$.
The electromagnetic environment enters Eq.~\eqref{tunnelHam} through 
fluctuating phase operators, $\theta_j$, which cause dephasing on the respective QD \cite{Devoret1990,Girvin1990,Nazarov,Altland}. 
For simplicity, these fluctuations are described by a single bosonic bath,
$H_{\rm env}=\sum_m E_m b^\dagger_m b^{}_m$, where $E_m>0$ and $\theta_j=\sum_m g_{j,m}(b_m+b_m^\dagger)$ with couplings $g_{j,m}$.  
The phases $\theta_j$ appear below only via the combination $\theta=\theta_1-\theta_2$, where we define      
a bath spectral density ${\cal J}(\omega)=\pi\sum_m( g_{1,m}-g_{2,m})^2 E_m^2 \delta(\omega-E_m)$.
We study the most relevant Ohmic case, see also Ref.~\cite{Munk2019},
\begin{equation}\label{Ohmic}
 {\cal J}(\omega) = \alpha \omega e^{-\omega/\omega_c},
\end{equation}
where $\alpha=\frac{e^2}{2h}{\rm Re} Z(\omega=0)$ is a dimensionless system-bath coupling and frequencies above the scale $\omega_c$ are suppressed. 
Here $Z(\omega)=[Z_0^{-1}(\omega) + i\omega C_0]^{-1}$ is the dynamical impedance of the environment, see Fig.~\ref{fig1}(b), 
and we study the regime $\alpha<1$.
 To avoid photon-assisted excitations of above-gap quasiparticles or higher-charge states on the islands,
we demand $\omega_c\ll E_C,\Delta$. Non-Ohmic environments \cite{Breuer2006}   can similarly be studied. However, for the sub-Ohmic case, the mapping to a Lindbladian master equation is problematic, while for the super-Ohmic case, dissipative gaps can become very small. 

\emph{Lindblad equation.---}We focus on the weakly driven regime defined by
\begin{equation}\label{basiccond}
  \tilde g_0\ll T\ll  \omega_0, \quad A\alt \tilde g_0,\quad \tilde g_0=t_0^2 t_{LR}/E_C^2,
\end{equation}
where the energy scale $\tilde g_0$ characterizes the relevant cotunneling processes (see below).  
The driving-induced rate for pumping electrons between QDs 1 and 2 is thus assumed small against  
cotunneling rates. The condition $\tilde g_0\ll T$ is needed to justify the Born-Markov approximation, while $T\ll \omega_0$ is required for the 
rotating wave approximation.  Both approximations are used for deriving the Lindblad equation.
Next we switch to the interaction picture with respect to $H_{\rm d}+H_{\rm env}$, and use third-order perturbation theory in the tunnel couplings to project the theory to the lowest-energy charge sector of each island. 
We then trace the von-Neumann equation over the bath and the QD degrees of freedom.
As a result,
the reduced density matrix,  $\rho_{\rm M}(t)$, describing the Majorana sector obeys the Lindblad  equation \cite{PRB} 
 \begin{equation} \label{LindbladMBQ}
    \partial_t\rho_{\rm M}(t)=-i[H_{\rm L},\rho_{\rm M}(t)]+
    \sum_{n=1,2}  \Gamma_n\mathcal{L}[K_n] \rho_{\rm M}(t),
\end{equation}
where the dissipator ${\cal L}$ acts on $\rho_{\rm M}$ according to \cite{Breuer2006}
$\mathcal{L}[K]\rho_{\rm M}=K\rho_{\rm M} K^\dagger -\frac{1}{2}\lbrace K^{\dagger} K,\rho_{\rm M} \rbrace$.
With $\lambda_{LR}\equiv t_{LR}/E_C \ll 1$ and the Pauli operators \eqref{PauliOp}, we obtain the two jump operators in Eq.~\eqref{LindbladMBQ} as
\begin{equation}\label{jumpK}
  K_1^{} =ie^{i\beta_2}|\lambda_{2,3R}| \left( e^{-i\beta_1}\frac{|\lambda_{1,1R}|}{\lambda_{LR}} X_R +i |\lambda_{1,3L}| Z_L Y_R\right)
\end{equation}
and $K_2=K_1^\dagger$, using the gauge choice $\lambda_{1,1R}=|\lambda_{1,1R}|e^{-i\beta_1}, \lambda_{1,3L}=|\lambda_{1,3L}|,$ and $\lambda_{2,3R}=|\lambda_{2,3R}|e^{-i\beta_2}$. 
The coherent  evolution in Eq.~\eqref{LindbladMBQ} is due to the Hamiltonian 
\begin{eqnarray}\label{hq}
  H_{\rm L} & = & 2p\tilde g_0 K_z + \sum_{n=1,2} h_n K_n^\dagger K_n, \\ \nonumber
  K_z^{} &=&\sin\beta_1|\lambda_{1,1R}\lambda_{1,3L}| Z_LZ_R.
\end{eqnarray}
For the spectral density in Eq.~\eqref{Ohmic}, assuming $\omega_c\gg \omega_0$, the dissipative transition rates $\Gamma_n$ and the Lamb shift parameters $h_n$ are 
given by 
\begin{eqnarray}\nonumber
\Gamma_1 &=& 2p \Gamma(1-2\alpha) \sin(2\pi \alpha) \left(\frac{\omega_0}{\omega_c}\right)^{2\alpha} \frac{\tilde g_0^2}{\omega_0}   ,\\
\Gamma_2 &=& \frac{(1-p) }{2p} e^{-\omega_0/T} \Gamma_1,\label{trans22} \\ \nonumber
h_1&=& \frac12 \cot(2\pi \alpha) \Gamma_1 ,\quad h_2=\frac{(1-p)}{2p}e^{-\omega_0/T} h_1,
\end{eqnarray}
where $p\approx A/\omega_0$ is the steady-state occupation probability of the energetically high-lying QD 2, and $\Gamma(z)$ denotes the Gamma function.

At low temperatures, $T\ll \omega_0$, the ratios $\Gamma_{2}/\Gamma_1$ and  $h_2/h_1$ are exponentially small, and therefore only the jump operator $K_1$ 
is important. 
This operator can be traced back to \emph{unidirectional} cotunneling transitions, where an electron is transferred from QD 2
 to QD 1 by cotunneling through the double-box setup. In the steady state, a weak drive amplitude $A$ is  then
responsible for pumping the dot electron back (from QD $1\to 2$) via the driven tunnel link.  
We note that the parameters $A, \omega_{0}, \alpha, \omega_c,$ and $\tilde g_0$
only affect the rates $\Gamma_n$ and Lamb shifts $h_n$, which in turn determine the speed of approach towards the dark space. 
The dark space itself, however, will be determined by the choice of the jump operator $K_1$, which can be 
engineered by tuning the `state design parameters' $\lambda_{j,\nu\kappa}$, see Eq.~\eqref{jumpK}.  These parameters
 can be adjusted via gate voltages. 
The ability to design jump operators via unidirectional cotunneling processes in such a manner
is rooted in the nonlocal Majorana representation of the Pauli operators in Eq.~\eqref{PauliOp}, and thus in the underlying topological nature of our DD system.

\emph{Dissipative maps.---}The key idea of our DD protocols is to choose the
state design parameters $\lambda_{j,\nu\kappa}$
such that $K_1$ implements a selected dissipative map \cite{Albert2014,Albert2016},
which in turn directly drives $\rho_{\rm M}(t)$ to the desired dark space. Below  we use the four Bell states,
$|\psi_\pm\rangle=(|00\rangle\pm |11\rangle)/\sqrt2$ and $|\phi_\pm\rangle=(|01\rangle\pm |10\rangle)/\sqrt2$, which are eigenstates of both $Z_LZ_R=\pm 1$ and $X_LX_R=\pm 1$ and span the Hilbert space of the two qubits in Eq.~\eqref{PauliOp}. 
We then define the dissipative maps $\hat E_{1,\pm} = (\mathbb{1}\pm Z_L Z_R) X_R,$ see Ref.~\cite{Barreiro2011}.
In the Lindblad equation, a dissipative term $\propto {\cal L}[\hat E_{1,-}]\rho_{\rm M}$ will map even-parity ($Z_LZ_R=+1$) states onto the respective odd-parity states, e.g.,
$\hat E_{1,-}|\psi_\pm\rangle=|\phi_{\pm}\rangle$. 
In contrast, odd-parity states do not evolve in time, $\hat E_{1,-}|\phi_\pm\rangle=0$, and thus represent steady-state solutions.  (Similarly, $\hat E_{1,+}$ maps odd-parity to even-parity states.)
As is shown next, under the dissipative map $\hat E_{1,-}$, the system can then be driven into the degenerate odd-parity sector spanned by $|\phi_{+}\rangle$ and $|\phi_-\rangle$.
 
\emph{Dark space stabilization.---}We now choose the parameters in Eq.~\eqref{tunnelHam} as
\begin{equation}\label{stabil1}
\beta_1 = \pi,\quad |\lambda_{1,1R}|=\lambda_{LR}|\lambda_{1,3L}|,
\end{equation}
with arbitrary $|\lambda_{2,3R}|$ and $\beta_2$.  Inserting Eq.~\eqref{stabil1} into Eq.~\eqref{jumpK} shows that    
 $K_1\propto \hat E_{1,-}$.
Moreover, from Eq.~\eqref{hq} we obtain $H_{\rm L}\propto Z_LZ_R$.  
Since the dissipator in Eq.~\eqref{LindbladMBQ} only involves $K_1\propto \hat E_{1,-}$ at $T\ll \omega_0$,
 the desired dissipative map can be realized without obstruction from the Hamiltonian dynamics.  
For this case, we can identify four conserved quantities, cf.~Ref.~\cite{Albert2014},
\begin{equation}
    C_{1,\pm} = \frac12(\mathbb{1}\pm Z_L),\quad
    C_{2,\pm} =\frac12(X_L \pm iY_L)X_R.
 \label{conservedquantities}
\end{equation}
The basis of the matrix Hilbert space corresponding to the dark space \cite{Albert2014} then follows as
\begin{eqnarray}
    M_{1,\pm} &=& \frac{1}{4}(\mathbb{1}\pm Z_L)(\mathbb{1}\mp Z_R),\nonumber\\
    M_{2,\pm} &=& \frac{1}{4}(X_L\pm iY_L)(X_R\mp iY_R).\label{Mbasis}
\end{eqnarray} 
The above DD protocol thus stabilizes a  degenerate dark space of
dimension $D=4$ in the terminology of Refs.~\cite{Albert2014,Albert2016}, which in turn coincides with the dark space dimension of a stabilized qubit space.
The Pauli operators $(X_D,Y_D,Z_D)$ for the resulting \emph{dark Majorana qubit} can be chosen as
\begin{eqnarray}\label{darkPauli}
    X_D&=&X_LX_R= -\gamma_1^L\gamma_3^L\gamma_1^R\gamma_3^R,\\ \nonumber
    Y_D&=&Y_LX_R=\gamma_2^L\gamma_3^L\gamma_1^R\gamma_3^R,\quad Z_D=Z_L=i\gamma^L_1\gamma_2^L.
\end{eqnarray}
The DD qubit encoding \eqref{darkPauli} is essential for fault tolerance, comparable to the formation of logical vs physical qubits in surface codes \cite{Fowler2012}. 
In our case, the DD protocol adds an extra protection layer on top of the topological protection of a native Majorana qubit. In particular, pure states
will thereby be stabilized for indefinite time \cite{footnotenew}.

\begin{figure}[t]
\begin{centering}
\includegraphics[width=0.9\columnwidth]{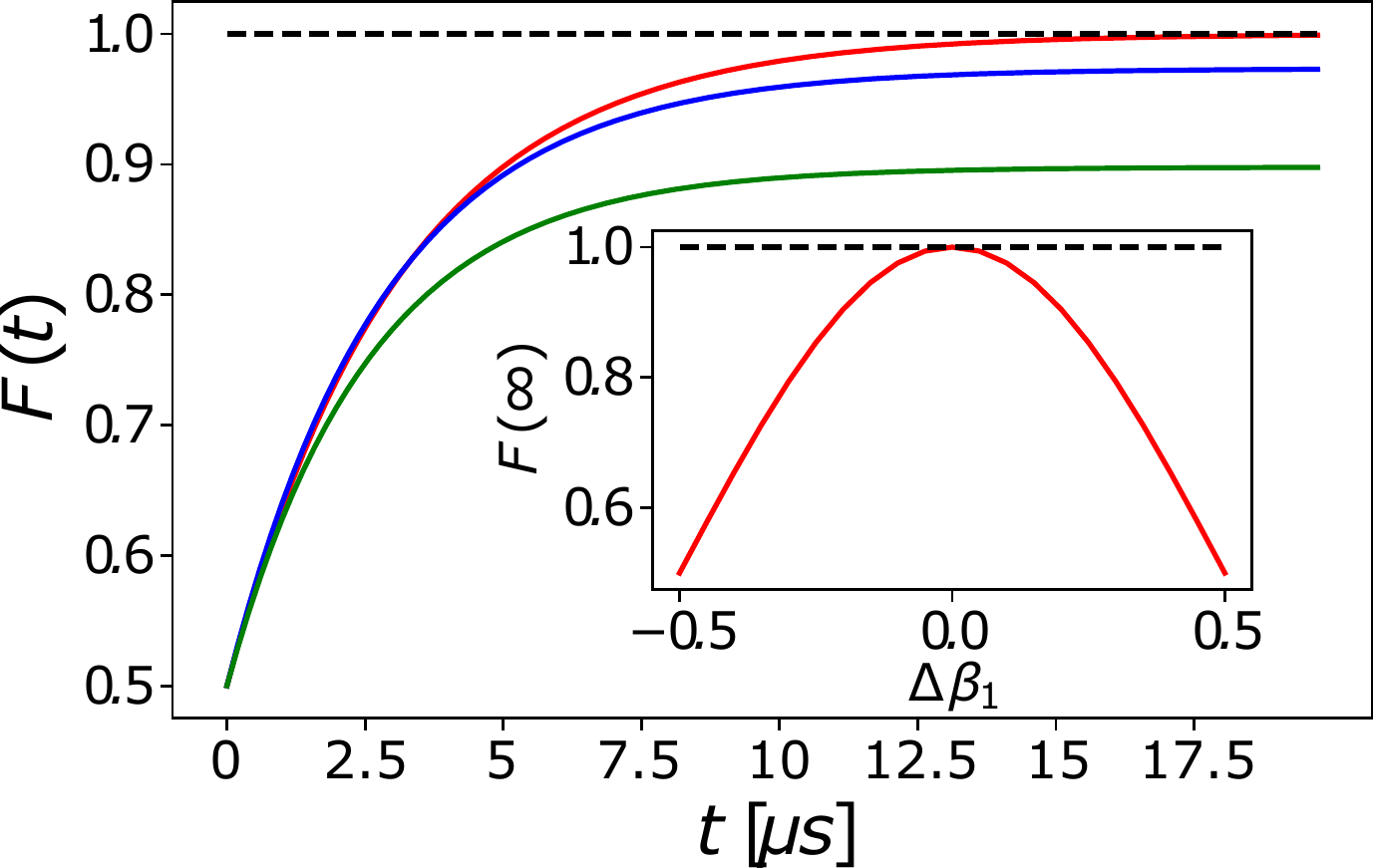}
\end{centering}
\caption{Fidelity for approaching the dark space starting from a maximally mixed initial state.
Main panel: Fidelity vs time, with $E_C = 1$~meV, $\tilde g_0/E_C=10^{-4}, T/\tilde g_0=2, \omega_0/\tilde g_0=200, \omega_c/\tilde g_0 = 10^3, \alpha=0.99, p(A) = 0.01$, and $|\lambda_{2,3L}|=1$.  
The red curve is for ideal state design parameters, see Eq.~\eqref{stabil1}, with $|\lambda_{1,1R}|=1$. 
The blue (green) curve is for parameters with $10\%$ ($20\%$) deviation from their 
respective ideal values.  Inset: Asymptotic ($t\to \infty$) fidelity vs percentage
deviation $\Delta\beta_1$ from $\beta_1=\pi$, with otherwise ideal parameters.} 
\label{fig2}
\end{figure}

\emph{Approaching the dark space.---}Starting from an arbitrary initial state $\rho_{\rm M}(0)$, 
we monitor the approach towards a pure target state $|\Psi\rangle$ in terms of the
fidelity, $F(t)={\rm tr} \left[  |\Psi\rangle\langle \Psi|\rho_{\rm M}(t)\right],$ 
where $\rho_{\rm M}(t)$ is the solution of Eq.~\eqref{LindbladMBQ}.
During the time evolution, all symmetry properties of the initial state other than parity remain preserved. For example, starting with $\rho_{\rm M}(0)=|\psi_+\rangle\langle\psi_+|$, since $X_LX_R=+1$ is kept as one approaches the target state, one finds $|\Psi\rangle=|\phi_+\rangle$ within the dark space.
In Fig.~\ref{fig2}, we show the fidelity obtained by numerical integration of Eq.~\eqref{LindbladMBQ} for a maximally mixed initial state $\rho_{\rm M}(0)=\frac{1}{4} \mathbb{1}\otimes \mathbb{1}$, where the corresponding target state is $|\Psi\rangle=(|\phi_+\rangle+|\phi_-\rangle)/\sqrt2$.
We note that if the initial state is not precisely known, one can first stabilize an arbitrary state inside the dark space, and subsequently drag that state towards the desired target state using the method described below.  A convenient way to initialize the dark qubit is to employ the tunnel couplings to a third QD \cite{PRB}.
Figure \ref{fig2} demonstrates that the dark-space fidelity is extremely robust against variations of the stabilization parameters $\lambda_{j,\nu\kappa}$.  Even when allowing for `errors' of 20$\%$ in \emph{all} these parameters, the fidelity is still $F\approx 0.9$. 
The time scale for approaching the steady state is given by the inverse of the \emph{dissipative gap} $\Delta_{\rm diss}$, which is the smallest real part of the non-zero eigenvalues of the Lindbladian dissipator.
For the above DD protocol, we obtain $\Delta_{\rm diss}\simeq |4\lambda_{1,3L}\lambda_{2,3R}|^2\sum_n\Gamma_n$, resulting in $\Delta_{\rm diss}^{-1}\approx 3~\mu$s for the parameters in Fig.~\ref{fig2}. 
Finally, methods for readout of the target state can be formulated as for the native Majorana qubit \cite{Plugge2017,Karzig2017,PRB}.

\begin{figure}[t]
\begin{centering}
\includegraphics[width=0.9\columnwidth]{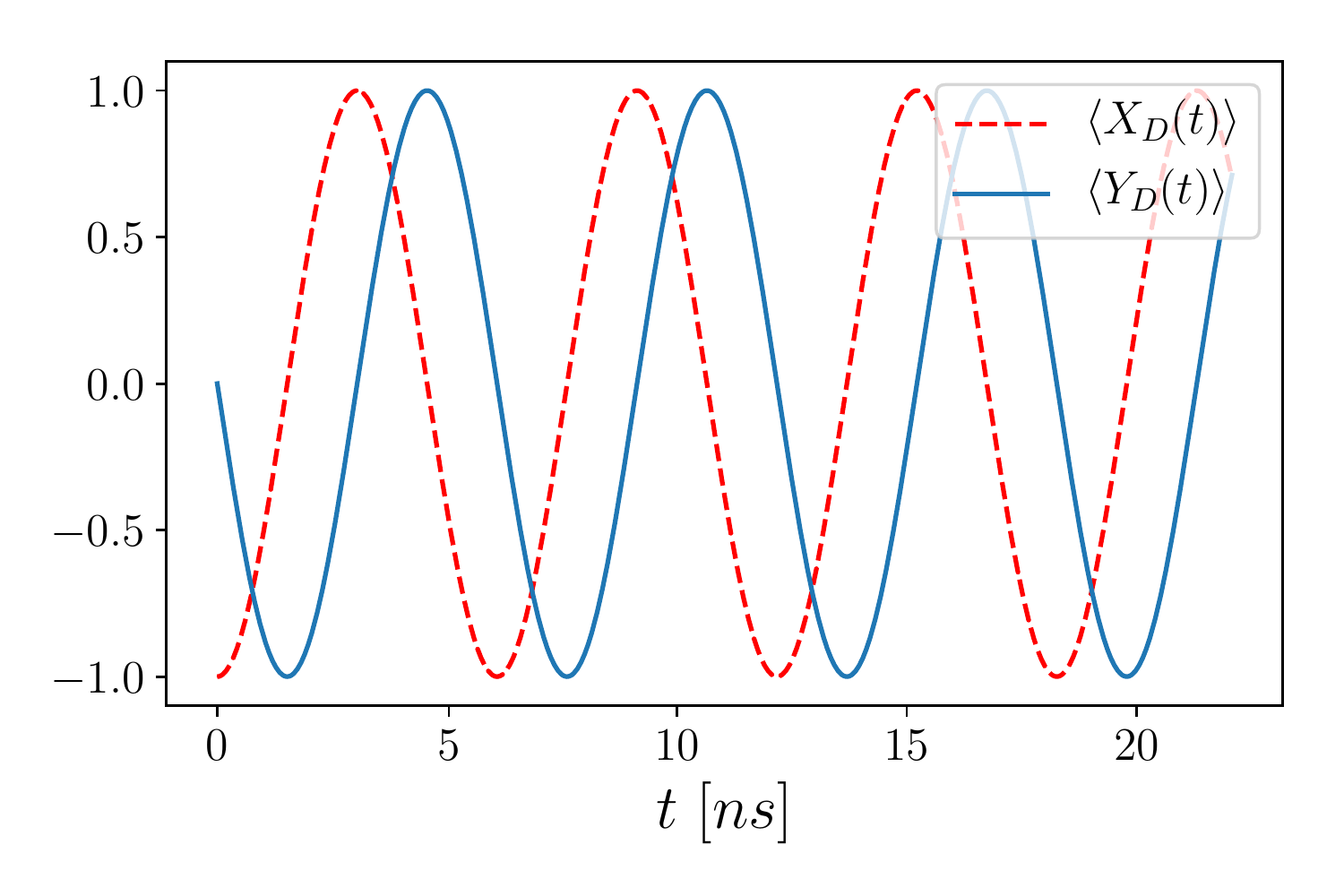}
\end{centering}
\caption{State manipulation by a $Z_L$-drive (see main text) with $A_{Z}/E_C=10^{-4}$.
We use the state design parameters in Eq.~\eqref{stabil1} and other parameters as in Fig.~\ref{fig2}.
The dynamics of the expectation values of the Pauli operators (\ref{darkPauli}) 
reveals oscillatory qubit coherences.  At all times, we numerically find $\langle Z_D\rangle=0$ and, of course,
$\langle Z_LZ_R\rangle=-1$ (odd parity).}
\label{fig3}
\end{figure}

\emph{State manipulation.---}We next discuss a general manipulation protocol moving an initial pure state in the dark space to an arbitrary final state in the dark space.
We adiabatically switch on a perturbation breaking at least one conservation law in Eq.~\eqref{conservedquantities}. The perturbation breaks the qubit degeneracy during the protocol but once the perturbation is switched off, the degenerate dark space is fully stabilized again. 
The main challenges are to avoid coupling the dark space to other Hilbert space sectors that are not part of the decoherence-free subspace, and to preserve the purity
of the state.  In particular, the drive should not connect odd- and even-parity sectors.
It is convenient to break two conserved quantities at any given time, leaving a two-fold degeneracy. 
The simplest protocol employs a `$Z_L$-drive' realized by 
coupling $\gamma_1^L$ and $\gamma_2^L$, see Eq.~\eqref{darkPauli}.   We thus add a term  
$H_{Z} = iA_{Z}(t)\gamma_1^L\gamma_2^L=A_{Z}(t) Z_L.$
The hybridization energy $A_{Z}(t)$ can be adiabatically changed using a gate-tunable tunnel link.
 $H_{Z}$ commutes with $Z_LZ_R$ and thus conserves parity.  The evolution generated by $H_{Z}$ therefore automatically remains in the odd-parity sector. Since $[H_{Z},C_{2}]\neq 0$ and 
$[H_{Z}, C_{3}]\neq 0$, see Eq.~\eqref{conservedquantities}, dark state coherences now depend on time. 
This is confirmed by our numerical results for constant $A_Z$ in Fig.~\ref{fig3}, where we start from $\rho_{\rm M}(0)=|\phi_-\rangle\langle\phi_-|$ and find 
oscillations in the real part, $\langle X_D(t)\rangle = \langle X_LX_R\rangle(t)$,
and the imaginary part,  $\langle Y_D(t)\rangle=\langle Y_LX_R\rangle (t)$, of the coherences.  
In the Bloch vector representation, the dark state periodically rotates in the $xy$-plane with
oscillation period $A_Z^{-1}$, where $A_Z^{-1}\approx 6$~ns in Fig.~\ref{fig3}.
For a general adiabatic protocol $A_Z(t)$, it stands to reason that an arbitrary final state inside the dark space can be reached. 

\emph{Conclusions.---}We have introduced a DD Majorana platform for stabilizing a degenerate dark space which offers several key advantages.  In particular, the spatial disentanglement of drive and dissipation processes rooted in the topological protection of MBSs allows for remarkably high levels of robustness. Future work should address the fidelity and purity during state manipulations and the high-dimensional dark spaces in DD systems with many coupled boxes.

\begin{acknowledgments} 
We thank A. Altland,  S. Diehl, and K. Snizhko for helpful discussions.
This project has been funded by the Deutsche Forschungsgemeinschaft (DFG,
German Research Foundation) under Grant No.~ 277101999, TRR 183 (project
C01), under Germany's Excellence Strategy - Cluster of Excellence Matter
and Light for Quantum Computing (ML4Q) EXC 2004/1 - 390534769, and under Grant No.~EG 96/13-1. In addition, we acknowledge funding
 by the Israel Science Foundation.  
\end{acknowledgments}

\end{document}